# More on the low variance circles in CMB sky

V.G.Gurzadyan[1], R.Penrose[2]

1. Yerevan Physics Institute and Yerevan State University, Yerevan, 0036, Armenia
2. Mathematical Institute, 24-29 St Giles', Oxford OX1 3LB, U.K.

Two groups [3,4] have confirmed the results of our paper concerning the actual existence of low variance circles in the cosmic microwave background (CMB) sky. They also point out that the effect does not contradict the LCDM model — a matter which is not in dispute. We point out two discrepancies between their treatment and ours, however, one technical, the other having to do with the very understanding of what constitutes a Gaussian random signal. Both groups simulate maps using the CMB power spectrum for LCDM, while we simulate a pure Gaussian sky plus the WMAP's noise, which points out the contradiction with a common statement [3] that 'CMB signal is random noise of Gaussian nature'. For as it was shown in [5], the random component is a minor one in the CMB signal, namely, about 0.2. Accordingly, the circles we saw are a real structure of the CMB sky and they are not of a *random* Gaussian nature. Although the structures studied certainly cannot contradict the power spectrum, which is well fitted by LCDM model, we particularly emphasize that the low variance circles occur in *concentric families*, and this key fact cannot be explained as a purely random effect. It is, however a clear prediction of conformal cyclic cosmology.

In our recent paper we have shown that the cosmic microwave background (CMB) sky has a characteristic property of possessing low variance concentric circles [1]. Such structures are expected on the basis of conformal cyclic cosmology (CCC) [2].

In the recent preprints [3,4] the authors aim to repeat our analysis based on the Wilkinson Microwave Anisotropy Probe (WMAP) 7 year data. They confirm the temperature variance structure obtained in our paper almost exactly, in particular directions given by us. It may be pointed out, however, that owing to a difference in the mean variance in different regions, it is not particularly informative to provide all them on a single plot as is done in [3] (see their Fig.1).

There are two radical differences, between our analysis and theirs. These are interrelated, one being a technical point and the other having to do with a matter of principle, concerning general understanding.

First, there is difference between our treatments of the Gaussian significance of the low variance circles. In [3,4] the authors simulate the CMB maps, as is commonly done, using the power spectrum parameters for LCDM plus the WMAP noise. As a result, they find standard deviation around 5μK, and hence conclude that there is a lower significance to the circles, e.g. around 3σ at 15μK depth, than that which we have found, where we adopted a more appropriate procedure, namely to simulate an *isotropic* Gaussian signal, to see whether such circles are result of random fluctuations. Below, in addition what is given in [1], we represent the standard deviations for 3 types of simulated Gaussian maps: Gaussian map simulated for parameters of (1) W-band (of all horns); (2) corresponding foreground reduced; (3) foreground reduced plus the WMAP's noise:

| Source | Sigma (µK) |
|---|---|
| Simulated Gauss W | 2.5 |
| Simulated Gauss W FR | 2.2 |
| Simulated Gauss W FR + WMAP's noise | 2.4 |

From this, we see, as is also correctly mentioned in [3,4], that the WMAP signal is indeed dominated by a cosmological signal rather than by an instrumental noise. However, we see that the *sigma* in the simulated data is only half the one claimed in [3] for LCDM, hence the corresponding increase in the Gaussian significance of the signal that we find in the actual CMB data.

Second, the claim in [3] that the 'CMB signal behaves like a random noise of Gaussian nature', which seems to be a common misconception, is invalid. As it is shown in [5], solving an inverse problem by means of Kolmogorov's theorem and von Neumann's method, the random component is a relatively minor one, of around 20% of the overall CMB signal. This fact undermines the above-mentioned statement of [3], with its corresponding interpretation of the analysis.

Thus, we have shown the particular reason for the discrepancy in the estimation of the Gaussian significance of the low variance circles claimed by [3,4] and conclude that (a) the low-variance circles are real structures in the CMB sky and (b) that by their considerable significance they are not due to *random* Gaussian fluctuations.

Our final point is a key issue that appears not to be taken seriously into account by [3,4]. What we find are *families* of concentric low-variance circles (as predicted by the cosmological scheme of [2]), not just individual low-variance circles. The probability of finding such families is clearly immensely smaller, for Gaussian random fluctuations, than finding the same number of unrelated individual low-variance circles. The simulations obtained by [3], including those using equilateral triangles rather than circles, show no indication of the low-variance instances occurring in concentric families, which is what we indeed find with the circles.

In sum, there are structures in CMB which, although obviously not contradicting the LCDM model, are certainly not explained by it. Nevertheless, these structures are clear predictions of CCC [2].